\documentclass[preprint,prl,aps,amsmath,amssymb,superscriptaddress,floatfix]{revtex4-1}
\usepackage{latexsym}        % to get LASY symbols
\usepackage{graphicx}        % to insert PostScript figures
\usepackage{rotating}        % for sideways tables/figures
\usepackage{epstopdf}

\usepackage{dcolumn}% Align table columns on decimal point
\usepackage{bm}% bold math

\begin{document}

\preprint{}

\title{Coordination of the energy and temperature scales of pairing across the doping phase diagram of Bi$_2$Sr$_2$CaCu$_2$O$_{8+\delta}$}%


\author{T. J. Reber}
\author{S. Parham}
\affiliation{Dept. of Physics, University of Colorado, Boulder, 80309-0390, USA
}%
\author{N. C. Plumb}
\altaffiliation{Now at Swiss Light Source, Paul Scherrer Institut, CH-5232 Villigen PSI, Switzerland
}%
\affiliation{Dept. of Physics, University of Colorado, Boulder, 80309-0390, USA
}%

\author{Y. Cao}
\author{H. Li}
\author{Z. Sun}
\author{Q. Wang}
\affiliation{Dept. of Physics, University of Colorado, Boulder, 80309-0390, USA
}%

\author{H. Iwasawa}
\affiliation{ Hiroshima Synchrotron Radiation Center, Hiroshima University, Higashi-Hiroshima, Hiroshima 739-0046, Japan
}%

\author{J. S. Wen}
\author{Z. J. Xu}
\author{G. Gu}
\affiliation{ Condensed Matter Physics and Materials Science Department, Brookhaven National Labs, Upton, New York,11973 USA
}%

\author{S. Ono}
\affiliation{RIKEN Advanced Science Institute, 2-1,Hirosawa, Wako, Saitama 351-0198, Japan}

\author{H. Berger}
\affiliation{Ecole Polytechnique Federale de Lausanne, CH-1015 Lausanne, Switzerland}

\author{Y. Yoshida}
\author{H. Eisaki}
\author{Y. Aiura}
\affiliation{ AIST Tsukuba Central 2, 1-1-1 Umezono, Tsukuba, Ibaraki 3058568, Japan
}%
\author{G. B. Arnold}
\author{D. S. Dessau}
\affiliation{Dept. of Physics, University of Colorado, Boulder, 80309-0390, USA
}%

\date{\today}% It is always \today, today,

\begin{abstract}
Using a new variant of photoelectron spectroscopy, we measure the homogeneous near-nodal pairing ($\Delta$) and pair-breaking self-energy ($\Gamma_S$) processes for a wide range of doping levels of Bi$_2$Sr$_2$CaCu$_2$O$_{8+\delta}$. For all samples we find that the pairing extends above the superconducting transition T$_c$ to a scale T$_{Pair}$ that is distinct from the antinodal pseudogap scale T$^*$ and near but slightly above T$_c$. We find that $\Delta$ and T$_{Pair}$ are related with a strong coupling ratio 2$\Delta$ /k$_B$T$_{Pair}\approx6$ across the entire doping phase diagram, i.e. independent of the effects of antinodal pseudogaps or charge-density waves.
\end{abstract}

\maketitle

Much effort has been put forth to understand the doping trends across the phase diagram of the cuprate superconductors, as this represents one of our best avenues to understand the d-wave superconductivity and other exotic physics of these materials. Key among the effects still to be understood are the behavior of the superconductive pairing and related superconductive transition, the pseudogap phase\cite{TimuskStatt} and potentially related charge density waves \cite{Ghiringhelli,Comin}, and the ``strange-metal" non-Fermi liquid scattering rates \cite{HusseyReview}.  Associated with each of these are temperature scales ($T_c, T_{pair}, T^*$, etc) as well as energy scales, with an expectation that the energy and temperature scales are connected in a straightforward way. In the simplest mean-field pictures these are directly related (e.g $2\Delta/k_BT_c=3.54$ for the weak-coupling BCS superconductor\cite{Tinkham}), though very strong fluctuations or competition between two or more phases may break any natural relations. Searching for such energy/temperature relations is thus a critical method for pinpointing the predominant physical principles. Our overall failure at finding such relations is thus directly correlated with our generally poor understanding of the physics of the cuprates.

The most fundamental energy/temperature relation of a superconductor is the ratio of the pairing energy to transition temperature, which has the ratio $2\Delta/k_B*T_c$=3.54 for the weak-coupling BCS superconductor\cite{Tinkham}, with ratios up to a factor of two larger for the so-called strong-coupling superconductors.  While these ratios are generally agreed to be large for the cuprate superconductors, no universal relation as a function of doping has yet been uncovered. The basic phenomenology indicates that while $T_c$ has a peak in the middle of the phase diagram, the gap grows continuously towards the underdoping regime \cite{CampuzanoUDPseudogap,Huefner2Gap} so that the ratio $2\Delta/(k_B T_c)$ grows dramatically in the underdoped regime. This has been rationalized in a number of ways, including that phase fluctuations dramatically suppress $T_c$ even though the pairing strength is large \cite{EmeryKivelson}; that the gap principally studied in these measurements is a pseudogap that is unrelated to the superconductivity \cite{Huefner2Gap} or that a competing effect reduces $T_c$ even though the pairing strength is large \cite{NormanFriendorFoe}.

Here we show that by focusing on the near-nodal gaps that are uncontaminated by competing pseudogaps and by utilizing the new TDoS method\cite{ReberArcs} of analyzing ARPES (angle-resolved photoemission spectroscopy) data which greatly improves the accuracy of gap measurements, we in fact find a clear relation between the pairing energy and temperature scales across a very large part of the doping phase diagram, with strong implications as will be discussed later.

The strong momentum dependence of the gap energy in a d-wave superconductor makes ARPES a uniquely powerful tool for the study of these gaps\cite{DamascelliRMP,DessauShenReview}. However, ARPES efforts to measure the pairing in the cuprates have been hampered by a) relatively limited energy resolution, forcing the preponderance of efforts to focus on the antinodal regimes where the gaps are largest, b) the contamination of the antinodal superconductive pairing gap with pseudogaps, which is also maximal in antinodal regime of the Brillouin zone c) our lack of understanding of the ARPES lineshape of the cuprates \cite{ShenSchriefferPRL,CaseyLineShape}, reflecting the exotic electronic correlation effects of these materials as well as the “dirt” or heterogeneity effects observed in, for example, STM measurements \cite{DavisInhomogeneity}. These latter aspects mean that ARPES measurements of the gaps in the cuprates had been limited to approximate techniques such as leading-edge shifts \cite{ShenDwavePRL}, peak-separation of symmetrized spectra \cite{NormanNature}, and fits to approximate model functions\cite{NormanPRB}.

This situation changed recently with the introduction of ultra-resolution laser-ARPES \cite{KoralekLaserARPES} as well as the Tomographic (sliced) Density of States (TDoS) method of ARPES analysis \cite{ReberArcs,ReberPrepairing,ParhamImpurity}, which bypasses the unknowns of the ARPES lineshape and removes much of the effects of the heterogeneous ``dirt" effects that are for example observed in STM experiments \cite{PanInhomogeneity} (see supplementary materials). Combined with new methods for removing nonlinearities in the electron detection \cite{ReberNL}, a quantitative analysis of the small (but uncontaminated by pseudogap or CDW effects) near-nodal gaps using Dynes's simple and well-tested formula for a lifetime-broadened gapped density of states \cite{DynesFormula,Dynes2} is now possible. In addition to measuring the gaps more accurately than previous methods, the TDoS allows a separation of the homogeneous pair-breaking lifetimes or scattering rates $\Gamma_S$ (a self-energy effect) from the pairing strengths $\Delta$ in an ARPES measurement. Overall we are afforded a more accurate determination of the angle, temperature, and doping dependence of both $\Delta$ and $\Gamma_S$, bringing qualitatively new insights into the nature of the pairing and pair-breaking in the cuprates.

\begin{figure*}[htbp]
\includegraphics[width=140mm]{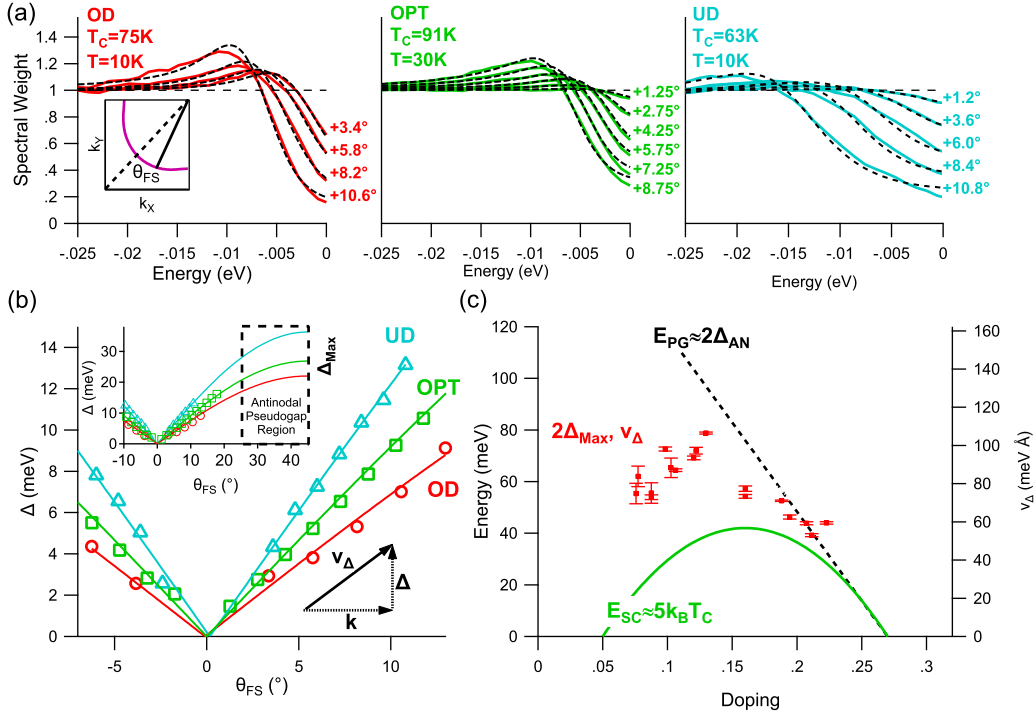}
\caption{\label{fig:Fig1}{ Doping dependence of the near-nodal pairing. (a) Angular dependence of the ARPES Tomographic Density of States (TDoS) as a function of angle $\theta_{FS}$ away from the node (see inset to upper panel) for three different dopings of Bi2212. The gap grows and deepens as $\Delta$ increases. Fits (black dashed) to the TDoS using equation 1 are shown. At the stated temperatures, the gap is fully formed as detailed in Fig. 2 (b) Extracted near-nodal $\Delta(\theta_{FS})$ as a function of doping from the fits of panel (a), including a d-wave extrapolation to the antinode at $\theta_{FS}=45^\circ$ (inset). The antinodal regime is contaminated by pseudogaps (boxed region, inset), which is why the near-nodal gap measurements are more accurate for determining the superconducting gap scale. We also indicate schematically how this data gives the scale $v_\Delta$ (see text for details). (c) $2\Delta_{Max}$ or equivalently $v_\Delta$ as a function of doping - data are compiled for 12 samples, 3 of which are shown in panels (a) and (b). These values follow neither the pseudogap energy scale, $E_{PG}$, nor the superconducting dome, $E_{SC}$, but rather trace an intermediate energy scale.}} \label{fig1}
\end{figure*}

In Figure 1a we show the angular or momentum evolution of the superconducting state TDoS in the near-nodal region for three dopings of Bi$_2$Sr$_2$CaCu$_2$O$_{8+\delta}$ (Bi2212). These curves show a depletion of spectra weight (the gap) very near zero energy (the Fermi energy, E$_F$) as well as the ``pile-up" of weight at slightly larger energies. As Fermi surface angle is measured from the zone corner, $\theta_{FS}$, increases away from the node, the gap grows, the depression of weight at E$_F$ deepens, and the pile-up of spectral weight moves to deeper energies. This rapid evolution of the spectral weight with momentum near the nodal point is only measurable with ARPES.  To quantify these results, we use the simplest model we can: the well-tested form first proposed by Dynes to explain the tunneling spectra of s-wave superconductors\cite{DynesFormula}:
\begin{equation} \label{eq:DynesFormula}
I_{Dynes}=Re\frac{\omega-i\Gamma_S}{\sqrt{(\omega-i\Gamma_S)^2-\Delta^2}}
\end{equation}

with fit results shown by the black dashed lines.  This formula describes a Bardeen-Cooper-Schreiffer (BCS)-like density of states with gap $\Delta$, modified by a broadening parameter $\Gamma_S$, which is generally described as the rate of pair-breaking or as the inverse of the pair lifetime.  It is equivalent to the inclusion of a self-energy term of the same magnitude into the Nambu-Gorkov single particle Green's function:
\begin{equation} \label{eq:NambuFormula}
G=\left( \begin{array}{cc}
\omega+i\Gamma_S-\epsilon_k & \Delta  \\
\Delta^* & \omega+i\Gamma_S+\epsilon_k
\end{array} \right)
\end{equation}
also indicating that it is equivalent to a single-particle scattering rate  $\Gamma_S$ in the electron self-energy term.  We note that the inclusion of this energy-independent homogeneous self-energy is the simplest first-order extension of the BCS model.  Including higher-order terms such as an energy-dependence to the self-energy or multiple types of scattering rates \cite{NormanPRB} give more free parameters than we find necessary for this present work.  Also, since minimal work has been yet done on the homogeneous self-energies with just the one broadening parameter, the present results are a necessary step forward.

We place the ``S" subscript on  $\Gamma$ to indicate that it is the superconductive self-energy or pair-breaking term, which is smaller than what is observed in the normal state. The single valued nature of $\Delta$ is appropriate for these fits since each TDoS spectrum is measured at an individual location along the Fermi surface, where the gap  $\Delta(k)$ is single valued.  This is an advantage over $k$-integrated spectroscopies such as tunneling, where specific $k$-dependent forms of the density of states and gap and pair-breaking functions must be assumed \cite{GomesPairFormation,RenGaps,AlldredgeGamma}.  The tunneling data also must make assumptions about the complicated pseudogap behavior at the antinode - effects which are absent in this near-nodal data.   As this simple extension of the BCS density of states captures the general behavior of the TDoS, we find that more complicated models are not necessary at this time, but future studies of the fine residual discrepancies may require additional terms or effects.

The extracted momentum dependence of $\Delta$ from these fits is shown in Figure 1b for the three dopings of Figure 1a. For all three dopings the gap grows linearly and symmetrically away from the node as expected for a d-wave superconductor. We fit this data with a simple d-wave form:
\begin{equation} \label{eq:DwaveFormula}
\Delta(\theta_{FS})=\Delta_{Max}|\sin{2\theta_{FS}}|
\end{equation}
A view of this extrapolation is shown in the inset to Figure 1b. Note that this is different from and arguably better than a direct measurement of the gap at the antinode, as we avoid any possible contribution from the separate (and likely competing) ``pseudogap" \cite{TanakaTwoGap,Kondo2Gaps,WiseCheckerboard}.  The three values of $\Delta_{Max}$, as well as those obtained similarly from many other samples, are compiled in Figure 1c. This type of measurement has occasionally been  characterized as ``gap velocity", v$_\Delta$.  Which has a simple relation in the small angle limit near the node:
\begin{equation} \label{eq:v_DeltaFormula}
\Delta(\theta_{FS})=v_\Delta k\approx v_\Delta (\sqrt{2}\frac{\pi}{a}-k_F)\theta_{FS}
\end{equation}
For the sake of ease of comparison to previous measurements, we include the corresponding values for v$_\Delta$ on the right axis of Fig. 1c.

Our measured gap values in Figure 1c are compared to the two energy scales compiled by H{\"u}fner et al.: $E_{SC}$ and $E_{PG}$ \cite{Huefner2Gap}.  $E_{PG}$ was primarily determined from the potentially competing anti-nodal gap magnitude, $\Delta_{AN}$ (STM, ARPES, etc.). Meanwhile, E$_{SC}$ was compiled from various measurements (Raman, INS) of bosonic modes that follow the superconducting dome. In this compilation, $E_{PG} \approx 2\Delta_{AN}$ while $E_{SC} \approx 5k_BT_c$. The superconducting gap energy measured here is clearly different from the two scales compiled by H{\"u}fner. While all three energy scales roughly agree in the over-doped region, they diverge around optimal doping. This divergence from E$_{PG}$ confirms that the near-nodal gap is distinct from the antinodal pseudogap physics \cite{TanakaTwoGap,Kondo2Gaps,WiseCheckerboard}. Qualitatively similar ARPES results for near-nodal gap magnitudes have been presented recently \cite{VishikTrisect}, though our data points to a general curvature of the gap scales with doping as opposed to the extremely flat dependence below optimal doping in this previous work, and as discussed next, our work also allows a connection between the temperature and energy dependence of the pairing scales as a function of doping.

\begin{figure*}[htbp]
\includegraphics[width=140mm]{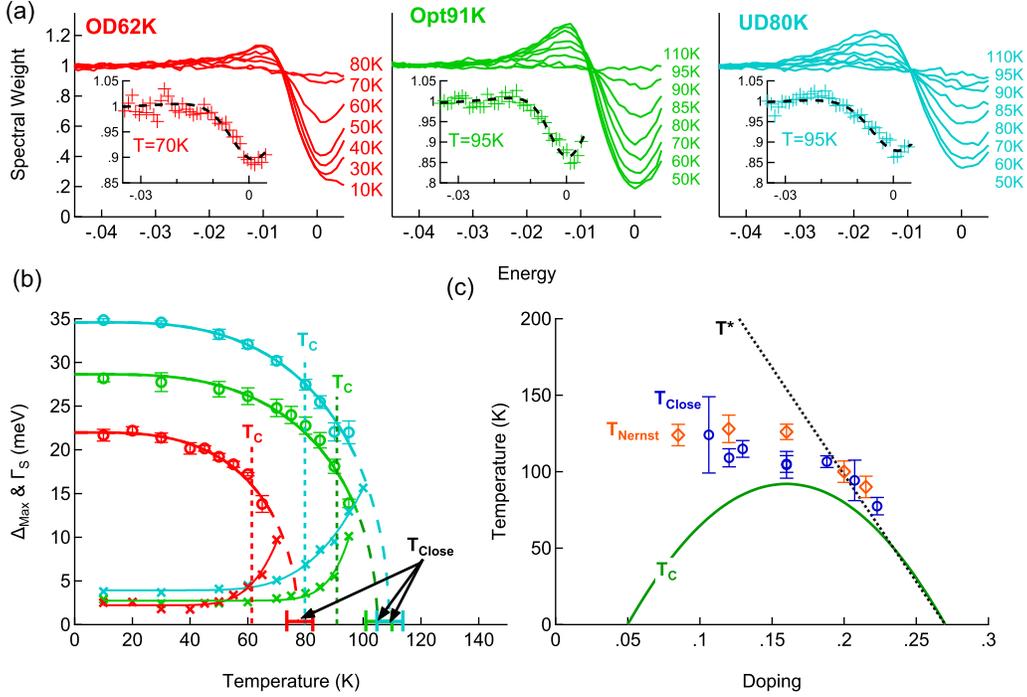}
\caption{\label{fig:Fig2}{Temperature dependence of pairing (a) Temperature dependence of the TDoS for three different dopings of Bi2212 at the near-nodal angle $\theta_{FS} = 12^\circ$. A gap above the T$_c$ is directly observed in all the samples (insets), indicating that pre-pairing is a universal phenomenon for all samples. (b) Extracted $\Delta_{Max}(T)$ and  $\Gamma_S(T)$ for the three samples, from fits of data including that of figures 1 and 2 to equation 1, with the error bars representing the returned uncertainty by the fits. We fit the temperature dependence of the gaps using equation 5, with the results shown by the dashed curves through the data points. These curves allow us to extrapolate to the temperature T$_{Close}$ at which the superconducting gaps close (go to zero). (c) The gap closing temperature T$_{Close}$ as a function of doping, compared to the reported T$_{Nernst}$ by Wang et al.\cite{WangPRB} which are both intermediate to T$_c$ and T$^*$.  The errors bars represent a combination of statistical and systematic error.}} \label{fig2}
\end{figure*}

	As we have shown previously \cite{ReberArcs,ReberPrepairing}, the traditional methods of determining the near nodal gap magnitude from ARPES fail near T$_c$ due to the large scale of $\Gamma_S$. Here we continue these studies by tracking the temperature dependence of the near-nodal gaps at many dopings. In Figure 2a, we show a representative temperature dependence of the TDoS for three different dopings at a single angle ($\theta_{FS}\approx 12^\circ$). Qualitatively, the constant location of the peak in any one panel shows that the gap size changes minimally over the full temperature range for each doping. Importantly, the gap is evident in the TDoS's even above T$_c$ for all dopings (insets of the top three panels), which is direct evidence for pre-pairing across the phase diagram.  To extract quantitative information of the temperature dependence of the gaps, we repeat the measurement procedure of figure 1 for a wide range of temperatures. For each temperature we extract a value of  $\Delta_{Max}$, which shrinks with increasing temperature.  The results are shown in Fig. 2b for three samples. For all dopings, we find that the superconducting gap smoothly evolves through and continues to exist above T$_c$ (vertical dashed lines). The smooth evolution of the gap shows that the near-nodal gap above T$_c$ has the same origin as the gap below T$_c$ - strong evidence of pre-superconductive pairing.\cite{WangNernst}

Figure 2b also shows details of the pair-breaking scattering rates  $\Gamma_S$, obtained from the same fits that returned the gap values.  $\Gamma_S$ has earlier been shown to be roughly constant as a function of Fermi surface angle \cite{ReberArcs,ReberPrepairing} so here we show the average of  $\Gamma_S(T)$ determined from all near nodal angles.  Consistent with our earlier results \cite{ReberArcs,ReberPrepairing} the scattering rates start at a low (few meV) value at low temperature, rising rapidly as the superconducting transition is approached.  This increase in the scattering rapidly ``fills in" the gap, as observed in all panels of figure 2a, even in the presence of a static gap magnitude. Such ``filling in" of the gap with increasing temperature is observed in many other experiments including optics \cite{BasovRMP}, all other tunneling measurements \cite{FischerRMP}, and thermodynamics \cite{TallonCoexistence} and appears to be a generic feature of the cuprates. The smooth lines of figure 2b are fits to the data using a BCS-like temperature dependent gap form \cite{DahmLifetimes}:
\begin{equation} \label{eq:BCSTFormula}
\Delta(T)=\Delta_{Max}\tanh\left(1.8\sqrt{\frac{T_{Close}}{T}-1}\right)
\end{equation}
where $\Delta_{Max}$ is the near-nodal gap maximum already discussed and T$_{Close}$ is the temperature at which the gap closes. The fits describe the temperature dependence quite well over the range of data fitted, and extrapolate to a T$_{Close}$ that occurs at an intermediate temperature that is neither T$_c$ nor T$^*$ (blue Fig 2c). This result is consistent with recent Nernst and diamagnetism experiments that have found evidence for fluctuating pairs above T$_c$ (orange Fig. 2c) \cite{WangBi2212,WangNernst,WangPRB}. A variety of other experiments have also recently indicated that fluctuating pairs exist up to this general temperature range \cite{GomesPairFormation,RenGaps,CorsonNature,GrbicPRB,HetelQC,DubrokaOpticsPrepairing}. The data shows that T$_{Close}$ is always greater than T$_c$, continuing to grow in the underdoped region where T$_c$ begins to shrink, confirming superconductive pre-pairing across the superconducting dome. We hereafter directly relate our fitted value T$_{Close}$ to the pairing onset T$_{Pair}$ (see the supplemental information for more discussion).

\begin{figure}[htbp]
\includegraphics[width=140mm]{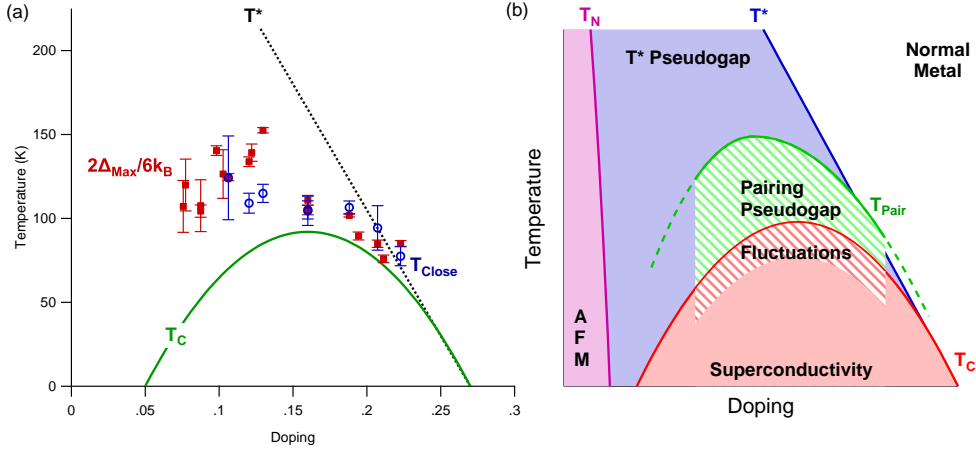}
\caption{\label{fig:Fig3}{ Updated doping phase diagram (a) A comparison of the temperature (blue circles) and energy scales (red crosses) of the pairing gap.  A scaling relation $2\Delta_{Max}/6k_BT_{Pair}$ is seen to hold across the entire phase diagram, where T$_{Pair}$ is defined as T$_{Close}$ from fig 2.  (b) A schematic view of the phase diagram.  In it we distinguish the activated pair-breaking fluctuation regime (hatched) which is centered around T$_c$ and extends up to T$_{Pair}$. Any gapping above T$_c$ has been called a pseudogap - here we should be clear to call this fluctuation regime the pre-superconductive pairing pseudogap region, which is present for all measured dopings. Additional, a separate ``T$^*$ pseudogap" extends up to T$^*$, which is principally present in underdoped samples and at the antinodal portion of the Brillouin zone.}} \label{fig3}
\end{figure}

Fig. 3a compares the doping dependence of the pairing energy scales, obtained from the gap energy $\Delta_{Max}(T\rightarrow0)$ and the temperature scale T$_{Pair}$. To convert the gap magnitude to a temperature we have scaled all gaps by the ratio $2\Delta_{Max}/6k_B$, which is observed to closely match T$_{Pair}$ (but not T$_c$) across the phase diagram. This is a clear deviation from BCS theory and the majority of expectations, in which it is T$_c$ and not T$_{Pair}$ that is correlated with the gap magnitude. A recent STM study showed a similar relation (with ratio 8) in the overdoped regime \cite{GomesPairFormation} though to our knowledge this is the first time such a scaling has been seen to span the full phase diagram, including the underdoped regime, where both the T$_c$ and the T$^*$ lines become well separated from the other scales.

Figure 3b generalizes the results of figure 3a, placing them into the wider context of the field\cite{BasovRMP,FischerRMP,TallonCoexistence,DahmLifetimes,WangNernst,WangBi2212,WangPRB}. In addition to showing the T$^*$ and T$_c$ lines with pseudogap and superconducting phases beneath them, we show the red hatched region of strong pair-breaking fluctuations, extending up to T$_{Pair}\gg$T$_c$, but also well below T$_c$ in the superconducting region (the regions in figure 2b where  $\Gamma_S$ starts increasing significantly with temperature - see supplemental materials for a more detailed discussion).

We have shown that the region between T$_{Pair}$ and T$_c$ has a pre-superconductive pairing-type pseudogap, but this must be different from any pseudogap physics that extends up to T$^*$. The T$^*$-pseudogap physics has long been known to exist principally near the antinodes \cite{LoeserScience,TimuskStatt}, and more recently has been discussed as being separate from and in fact competing with the superconductivity.  We thus explicitly label the phase diagram as having these two separate types of pseudogaps - a pre-superconductive pairing type extending up to T$_{Pair}>$T$_c$ and a ``T$^*$ pseudogap" extending up to T$^*$, which may for example be due to the competing phases. Note that while T$^*$ is usually well above T$_{Pair}$, the temperature ordering likely switches in the overdoped region, which is allowed since these scales originate from distinct processes.

Differing from previous phase diagrams, ours also shows a new scaling relation for the pairing gap. That T$_{Pair}$ rather than T$_c$ is correlated with the gap magnitude is fully consistent with pre-superconductive pairing ideas, implying that T$_{Pair}$ can nominally be associated with the mean field transition temperature. In many ways this should therefore be expected, unless one considers the great amount of work in the field focusing on the antinodal T* pseudogaps and CDW effects. Because these effects occur with greatly varying strength across the doping phase diagram (effectively occurring only in the underdoped region) it is therefore surprising how robust the pairing scaling relation is.  Our results show that $T_c$ deviates more from $T_{Pair}$ in the underdoped than overdoped regimes, and this may be due to the effect of the pseudogap. If this is the main effect of the T* pseudogap and/or CDW it would seem to indicate that these effects are rather small, i.e. second-order compared to the main effects of the near-nodal pairing energy $\Delta_{Max}$ and self-energy effect $\Gamma_S$.

Returning to the discussion of the pair-breaking self energy $\Gamma_S$, our results show this to be large, strongly temperature-dependent, and responsible for the ``filling in" of the gap (rather than closing of the gap), which is a phenomenology that has been observed in almost all spectroscopies of the gap in the cuprates, but rarely deconvolved as a specific effect.  Our results in fact suggest that the pair-breaking self energy $\Gamma_S$ plays a critical role in reducing the T$_c$ of these materials below the mean-field value T$_{Close}$.  This occurs as $\Delta$ decreases and the dynamic $\Gamma_S$ increases strongly with rising temperature.  We expect future work to focus closely on this important self-energy/pair-breaking effect, rather than just focusing on the pairing energy or pseudogap energies like the great majority of previous studies.

\begin{acknowledgments}
We thank A. Balatsky, E. Calleja, M. Hermele, I. Mazin, K. McElroy, L. Radzihovsky and Xiaoqing Zhou for valuable conversations and D. H. Lu and R. G. Moore for help at SSRL. SSRL is operated by the DOE, Office of Basic Energy Sciences. ARPES experiments at the Hiroshima Synchrotron Radiation Center were performed under proposal 09-A-48.  Funding for this research was provided by DOE Grant No. DE-FG02-03ER46066 (Colorado) and DE-AC02-98CH10886 (Brookhaven).
\end{acknowledgments}

%
%\bibliography{DeltaTPairbib}% Produces the bibliography via BibTeX.
\end{document}